\documentclass[showpacs,twocolumn]{revtex4}

\usepackage{graphicx}
\usepackage{dcolumn}
\usepackage{amsmath}
\makeatletter
\def\btt#1{\texttt{\@backslashchar#1}}
\DeclareRobustCommand\bblash{\btt{\@backslashchar}} \makeatother

\begin{document}

\title{Higher dimensional dust collapse with a cosmological constant}

\author{S. G.~Ghosh}\email{sgghosh@iucaa.ernet.in} \affiliation{BITS-Pilani, Dubai
Campus, P.B. 500022, Knowledge Village, Dubai - UAE}
\affiliation{Birla Institute of Technology and Science, Pilani -
333 031, INDIA}
\author{D. W.~Deshkar}
\affiliation{ Science College,
Congress Nagar, Nagpur 440 012, INDIA}%

\date{\today}

\begin{abstract}
The general solution of the Einstein equation for higher
dimensional (HD) spherically symmetric collapse of inhomogeneous
dust in presence of a cosmological term, i.e., exact interior
solutions of the Einstein field equations is presented for the HD
Tolman-Bondi metrics imbedded in a de Sitter background. The
solution is then matched to exterior HD Scwarschild-de Sitter.  A
brief discussion on the causal structure singularities and
horizons is provided. It turns out that the collapse proceed in
the same way as in the Minkowski background, i.e., the strong
curvature naked singularities form and that the higher dimensions
seem to favor black holes rather than naked singularities.
\end{abstract}


\maketitle
\section{Introduction}

That the results coming from the analysis of high redshift Type Ia
supernovae \cite{ag,sp,ss} indicate the Universe is accelerating.
This suggests  the possibility that a nonzero cosmological
constant ($\Lambda$) may dominate the total energy of our
Universe. The cosmological implications of the existence of a
$\Lambda$ term are enormous, concerning not only the evolution of
the Universe, but also structure formation and age problems. If
$\Lambda$ term must be restored to the Einstein equations,
surprises may turn up in other physical applications of Einstein's
equations as well. For example, Markovic and Shapiro \cite{ms}
generalized the Oppenheimer-Snyder model (which describes the
gravitational collapse of a spherical homogeneous dust ball
initially at rest in exterior vacuum to a Schwarzschild black
hole) taking into account the presence of a positive $\Lambda$.
They showed that $\Lambda$ may affect the onset of collapse and
decelerate the implosion initially. The results of the Markovic
and Shapiro were qualitatively generalized to the inhomogeneous
dust and degenerate cases by Lake \cite{kl} for both $\Lambda > 0$
and $\Lambda < 0$. It was shown explicitly by Cissoko {\it et al.}
\cite{cf} that the cosmological term slows down the collapse of
matter, limiting the size of black holes.

The Tolman-Bondi metric \cite{tb}, which is asymptotically flat,
has been extensively used to study the formation of naked
singularities in spherical collapse. It is seen that the
Tolman-Bondi metric admits both naked and covered singularities
depending upon the choice of initial data and there is a smooth
transition from one phase to the other
\cite{es,djd,jl,dj,jd1,rn,op,lz}. However, according to the cosmic
censorship conjecture (CCC) \cite{rp}, the singularities that
appear in gravitational collapse are always surrounded by an event
horizon. Moreover, according to the strong version of the CCC,
such singularities are not even locally naked, i.e., no
non-spacelike curve can emerge from such singularities (see
\cite{r1}, for reviews on the CCC). The study in the inhomogeneous
dust collapse with a positive $\Lambda$, from the viewpoint of
CCC, was examined in \cite{djcj,cf,sm,sg}. Deshingkar {\it et al.}
\cite{djcj} showed that the presence of a $\Lambda$ can cover a
part of the singularity spectrum which is visible in the
corresponding dust collapse models for the same initial data.
Whereas Wagh and Maharaj \cite{wm} showed that in spherically
symmetric radiation collapse (Vaidya collapse), the effect of
adding a positive $\Lambda$ does not radically alter the
description. Lemos \cite{jl1} arrived at the same conclusion for a
negative $\Lambda$. The result in both these cases are the same as
in the case of collapsing radiation in the Minkowskian background.
Therefore, at least, in the case of spherical radiation collapse,
the asymptotic flatness is not essential for the development of a
naked singularity.

While gravitational collapse has been originally studied in four
dimensions (4D), there have been several attempts, mainly
motivated by string theory, to study it in HD space-time
\cite{bsc,hd,gb,gd,gnd,kmy,gab,gj,usb}.  Since, current
experimental results involving tests of the inverse square law do
not rule out extra dimensions even as large as a tenth of a
millimeter. It is now important to consider the evolution of the
extra dimensions since the observed strength of the gravitational
force is directly dependent on the size of the extra dimensions.
As a consequence, there is a renewed interest towards
understanding of the general relativity in more than four
dimensions, as growing volume of recent literature indicates.  In
particular, several solutions to the Einstein equations of
localized sources in higher dimensions have been obtained in the
recent years \cite{rcm}.

In this paper, we shall study spherical inhomogeneous dust
collapse with  a positive $\Lambda$ in HD theory of gravity, and
present solutions in closed form. This is HD analogous of 4D
Tolman-Bondi-de Sitter solutions and for definiteness we shall
call it HD Tolman-Bondi-de Sitter solutions. Then, we show that HD
Tolman-Bondi-de Sitter admits strong curvature naked singularity.
However, the presence of a positive $\Lambda$ does not  radically
alter the established picture of Inhomogeneous dust collapse.

In the next section, we give exact HD spherically symmetric
solution of Einstein field equation for a collapsing inhomogeneous
dust with a cosmological constant $\Lambda$. This is followed by
junction conditions between a static and a non-static HD
spherically symmetric space-time in section III. The nature of
singularities of such a space-time, and the consequence of
cosmological constant $\Lambda$ is a subject of section V. This is
preceded by detailed analysis on apparent horizon in section IV.

We have used units which fix the speed of light and the
gravitational constant via $8\pi G = c^4 = 1$.
\section{Higher Dimensional Tolman-Bondi de Sitter space-times}
The standard 4D Tolman-Bondi solution \cite{tb} represents an
interior of a collapsing inhomogeneous dust sphere. The solution
we seek is - collapse of a spherical dust with a positive
$\Lambda$ in HD space-time. We choose a spherically symmetric
comoving metric in HD \cite{gb,bsc}, which has form
\begin{equation}
ds^2 =   dt^2 - e^{\lambda(t,r)} dr^2 - R(t,r)^2 d\Omega^2,
\label{eq:me}
\end{equation}
where
\begin{eqnarray}
d\Omega^2 = && \sum_{i=1}^{n} \left[ \prod_{j=1}^{i-1} \sin^2
\theta_j \right] d \theta_i^2 =
d \theta_1^2+ sin^2 \theta_1 d \theta_2^2\nonumber \\
&&+ sin^2 \theta_1 sin^2 \theta_2 d\theta_3^2+\;.\; .\;
.\;  \nonumber \\
&& + sin^2 \theta_1 sin^2 \theta_2 .\; .\; . sin^2 \theta_{n-1}
d\theta_n^2,\label{eq:ns }
\end{eqnarray}
is the metric on an $n$-sphere and $n=D-2$ (where $D$ is the total
number of dimensions), together with the stress-energy tensor for
dust:
\begin{equation}
T_{ab} =  \zeta(t,r) \delta_{a}^t \delta_{b}^t, \label{eq:emt}
\end{equation}
where $u_a = \delta_t^a$ is the $(n+2)$-dimensional velocity. The
coordinate $r$ is the co-moving radial coordinate, $t$ is the proper
time of freely falling shells, and $R$ is a function of $t$ and $r$
with $R>0$ and $\lambda$ is also a function of $t$ and $r$. With the
metric (\ref{eq:me}), the Einstein equations are
\begin{eqnarray}
G^t_t = && \frac{n(n-1)}{2 R^2} (e^{- \lambda} {R'}^2 - \dot{R}^2
-1) - \nonumber \\
&& \frac{n}{2}\frac{1}{R} ( \dot{R} \dot{\lambda} + e^{-\lambda} R'
\lambda') + n e^{-\lambda} \frac{R''}{R} = \zeta - \Lambda,
\label{eq:gtt}
\end{eqnarray}
\begin{equation}
G^r_r =  \frac{n(n-1)}{2 R^2} (e^{- \lambda} {R'}^2 - \dot{R}^2 -1)
- n \frac{\ddot{R}}{R} = - \Lambda,\label{eq:grr}
\end{equation}
\begin{eqnarray}
&& G^{\theta_1}_{ \theta_1} =  G^{\theta_2}_{\theta_2} =\; . \; . \;
. \; = G^{\theta_{n}}_{ \theta_{n}}  = \frac{(n-1)(n-2)}{2}
\frac{1}{R^2} \nonumber \\
&& \times (e^{- \lambda}{R'}^2 -  \dot{R}^2-1) -
\frac{n-1}{2}\frac{1}{R} (\dot{R} \dot{\lambda} + e^{- \lambda} R'
\lambda') - \nonumber \\
&& (n-1) \frac{1}{R}(\ddot{R} - e^{- \lambda} R'')-\frac{1}{2}
(\ddot{\lambda}+\frac{\lambda^2}{2}) = - \Lambda,\label{eq:gth}
\end{eqnarray}
\begin{equation}
G^t_r = \frac{n}{2R} (2 \dot{R}'- \dot{\lambda} R') = 0,
\label{eq:gtr}
\end{equation}
where the overdot and the prime denote the partial derivative with
respect to t and r, respectively. Integration of Eq.~(\ref{eq:gtr})
gives
\begin{equation}
e^{\lambda(t,r)} = \frac{{R'}^2}{1+f(r)}, \label{eq:el}
\end{equation}
which can be substituted into Eq.~(\ref{eq:grr}) to yield
\begin{equation}
\dot{R}^2 = \frac{\mathcal{M}(r)}{R^{n-1}} + \frac{2 \Lambda
R^2}{n(n + 1)} + f(r). \label{eq:fe}
\end{equation}
The functions  $\mathcal{M}(r)$ and  $f(r)$ are arbitrary and
referred to as  the mass and energy functions, respectively. Since
in the present discussion we are concerned with gravitational
collapse, we require that $\dot{R}(t,r) < 0$. The energy density
$\zeta(t,r)$ is calculated as
\begin{equation}
\zeta(t,r) = \frac{n \mathcal{M}'}{2 R^n R'},  \label{eq:edt}
\end{equation}
For physical reasons, one assumes that the energy density $\zeta(t,r)$ is
everywhere non-negative. The special case $f(r)=0$ corresponds to
the marginally bound case which is of interest to us in this paper.
Substituting Eqs.~(\ref{eq:el}) and ~(\ref{eq:fe}) into
Eq.~(\ref{eq:gtt}) yields
\begin{equation}
\mathcal{M}' = \frac{2}{n} \zeta R^n R'.\label{eq:m}
\end{equation}
Integrating Eq.~(\ref{eq:m}) leads to
\begin{equation}
\mathcal{M}(r) = \frac{2}{n} \int \zeta R^n dR, \label{eq:m1}
\end{equation}
where constant of integration is taken as zero since we want a
finite distribution of matter at the origin $r = 0$. The function
$\mathcal{M}(r)$ must be positive, because $\mathcal{M}(r) < 0 $
implies the existence of negative mass. This can be seen from the
mass function $\mathbf{M}(t,r)$, which is given by
\begin{eqnarray}
\mathbf{M}(t,r) & = & \frac{n-1}{2} R^{n-1} \left(1 - g^{ab} R_{,a} R_{,b} \right) \nonumber\\
&& = \frac{n-1}{2} R^{n-1} \left(1 - \frac{R{'}^2}{e^{\lambda}} +
\dot{R}^2 \right).\label{eq:m2}
\end{eqnarray}
Now Eqs.~(\ref{eq:el}), ~(\ref{eq:fe}) and ~(\ref{eq:m2}) implies
that
\begin{equation}
\mathbf{M}(t,r)  =  \frac{n-1}{2} \mathcal{M}(r) +
\frac{n-1}{n(n+1)} \Lambda R^{n+1}.
\end{equation}
The quantity $\mathcal{M}(r)$ can be interpreted as energy due to
the energy density $\zeta(t,r)$ given by Eq.~(\ref{eq:m1}), and
since it is measured in a comoving frame, $\mathcal{M}$ is only
$r$ dependent. Cissoko {\it et al}. \cite{cf} have derived
marginally bound $(f = 0)$ dust solution in the presence of
$\Lambda > 0$. Here we derive the analogous HD solutions.
Equation~(\ref{eq:fe}), for vanishing $\Lambda$, in 4D as well as
in HD, has three types of solutions, namely, hyperbolic, parabolic
and elliptic solutions depending on whether $f(r) > 0$, $f(r) = 0$
or $f(r) < 0$, respectively. The condition $f(r)= 0$ and $\Lambda
= 0$ is the marginally bound condition, limiting the situations
where the shell is bounded from those it is unbounded. In the
presence of a cosmological constant, the situation is more
complex, and $f(r) = 0$ leads to an unbounded shell. The
assumption allows for analytical solutions in closed form
\begin{equation}
R(t,r) = \left[\frac{n(n+1)\mathcal{M}}{2
\Lambda}\right]^{1/{n+1}} \; \sinh^{2/{n+1}}\alpha, \label{eq:R}
\end{equation}
\begin{eqnarray}
R'(t,r)&=& \left[\frac{n(n+1)\mathcal{M}}{2
\Lambda}\right]^{1/{n+1}}
\Big[\frac{\mathcal{M}'}{(n+1)\mathcal{M}} \; \sinh \alpha  \nonumber\\
& +& \sqrt{\frac{2\Lambda}{n(n+1)}} t_0' \; \cosh \alpha  \Big] \;
\sinh^{(1-n)/(1+n)}\alpha , \label{eq:RD}
\end{eqnarray}
where $\alpha = \alpha(t,r)$ has the form
\begin{equation}
\alpha (t,r) = \sqrt {\frac{(n+1) \Lambda}{2n}} [t_0(r) -
t],\label{eq:alp}
\end{equation}
where $t_{0}(r)$ is an arbitrary function of integration which
represents the proper time for the complete collapse of a shell
with coordinate $r$. It follows from the above that there is a
space-time singularity at $R=0$ and at $R'=0$.

It is easy to see that as $\Lambda \rightarrow 0$ the above
solution reduces to the HD Tolman-Bondi solutions \cite{gb}:
\begin{equation}
\lim_{\Lambda \rightarrow 0} R(t,r) =  \left[\frac{(n+1)^2
\mathcal{M}}{4} (t_0 - t)^2 \right]^{1/{n+1}}, \label{TBR}
\end{equation}
\begin{equation}
\lim_{\Lambda \rightarrow 0} R'(t,r) = \frac{\mathcal{M}' (t_0 - t)
+ 2\mathcal{M} t_0'}{\left[\frac{(n+1)^{n-1}}{4} \mathcal{M}^n (t_0
- t)^{n-1}\right]^{1/{n+1}}}. \label{TBRD}
\end{equation}
The standard 4D Tolmon-Bondi solution can be now recovered by
setting $n=2$. The three arbitrary functions $\mathcal{M}(r)$,
$f(r)$ and $t_{0}(r)$ completely specify the behavior of shells
with radius $r$. It is possible to make an arbitrary relabeling of
spherical dust shells by $r \rightarrow g(r)$, without loss of
generality, we fix the labeling by requiring that, on the
hypersurface $t = 0$, $r$ coincides with the radius
\begin{equation}
R(0,r) = r.
\end{equation}
This corresponds to the following choice of $t_{0}(r)$:
\begin{equation}
t_{0}(r) =
\sqrt{\frac{2n}{(n+1)\Lambda}}\;\sinh^{-1}\left[\sqrt{\frac{2\Lambda}{n(n+1)\mathcal{M}}}r^{(n+1)/2}\right].
\end{equation}
The central singularity occurs at $r = 0$, the corresponding time
being $t = t_{0}(0) = 0$. We denote by $\rho (r)$ the initial
density:
\begin{equation}
\rho(r) = \zeta(0,r) = \frac{n\mathcal{M}'}{2r^{n}} \Rightarrow
\mathcal{M}(r) = \frac{2}{n} \int \rho(r)r^{n} dr.
\end{equation}

\section{Junction conditions}
In order to study the gravitational collapse of a finite spherical
body we have to match the solution along the time like surface at
some $R = R_{\Sigma}$ to a suitable HD exterior. We consider a
spherical surface with its motion described by a time-like
(n+1)-surface $\Sigma$, which divides space-times into interior
and exterior manifolds $\mathcal{V}_I$ and $\mathcal{V}_E$.
According to the generalized Birkoff theorem \cite{bt} the vacuum
space-time outside is HD Schwarzschild-de Sitter space-time:
\begin{equation}
ds^2  =  F(Y) dT^2 - \frac{1}{F(Y)} dY^2 - Y^2 d\Omega ^2,
\label{eq:me1}
\end{equation}
where $F$ is a function of $Y$ given by
\begin{equation}
F(Y) = 1 - \frac{2M}{(n-1)Y^{n-1}} - \frac{2 \Lambda Y^2}{n(n+1)},
\end{equation}
and $M$ is a constant. In accordance with Darmois junction
condition, we have to demand when approaching $\Sigma$ in
$\mathcal{V}_I$ and $\mathcal{V}_E$
\begin{equation}
(ds_{-}^2)_{\Sigma} = (ds_{+}^2)_{\Sigma} = (ds^2)_{\Sigma},
\label{jc1}
\end{equation}
where the subscript ${\Sigma}$ means that the quantities are to be
evaluated on $\Sigma$ and let $K^{\pm}_{ij}$ is extrinsic
curvature to $\Sigma$, defined by
\begin{equation}
K^{\pm}_{ij} = - n_{\alpha}^{\pm} \frac{\partial^2
\chi^{\alpha}_{\pm}}{\partial \xi^i \partial \xi^j} -
n_{\alpha}^{\pm} \Gamma^{\alpha}_{\beta \gamma} \frac{\partial
\chi^{\beta}_{\pm} }{\partial \xi^i} \frac{\partial
\chi^{\gamma}_{\pm} }{\partial \xi^j}, \label{ec}
\end{equation}
and where $\Gamma^{\alpha}_{\beta \gamma}$ are Christoffel symbols,
$n^{\pm}_{\alpha}$ the unit normal vectors to $ \Sigma $,
$\chi^{\alpha}$ are the coordinates of the interior and exterior
space-time and $\xi^i$ are the coordinates that defines $\Sigma$.
The intrinsic metric on the hypersurface $r=r_{\Sigma}$ is given by
\begin{equation}
ds^2 =   dt^2 - R^2(r_{\Sigma}, t) d\Omega^2, \label{bm}
\end{equation}
with coordinates $\xi^a = (t, ~\theta_1, ~\theta_2, ~\theta_3, \; .
\; . \;, ~\theta_n)$. In this coordinate the surface $\Sigma$, being
the boundary of the matter distribution, will have the equation
\begin{equation}
r - r_{\Sigma} = 0, \label{se}
\end{equation}
where $ r_{\Sigma}$ is a constant. The first fundamental form of
$\Sigma$  can be written as $g_{ij}d\xi^i d\xi^j$.  Then the
exterior metric, on $\Sigma$, becomes:
\begin{equation}
ds_{\Sigma}^2  =  \left[ F(Y_{\Sigma}) - \frac{1}{F(Y_{\Sigma})}
\left(\frac{dY_{\Sigma}}{dT}\right)^2 \right] dT^2 - Y_{\Sigma}^2
d\Omega^2, \label{mhs}
\end{equation}
where we assume that the coefficient of $dT^2 >0$ so that $T$ is
time like coordinate.  From the first junction condition we obtain
\begin{eqnarray}
R(r_{\Sigma},t) = Y_{\Sigma}, \nonumber\\
\left[F(Y_{\Sigma}) - \frac{1}{F(Y_{\Sigma})}
\left(\frac{dY_{\Sigma}}{dT} \right)^2 \right]^{1/2} dT = dt.
\label{fjc}
\end{eqnarray}
The non-vanishing components of extrinsic curvature $K_{ij}^{\pm}$ of
$\Sigma$ can be calculated and the result is
\begin{eqnarray}
&& K^{+}_{t \; t} = \left[ \dot{Y} \ddot{T} - \dot{T} \ddot{Y} -
\frac{F}{2} \frac{dF}{dY} \dot{T}^3 + \frac{3}{2F} \frac{dF}{dY}
\dot{T} \dot{Y}^2 \right]_{\Sigma},  \label{ecia}\\
&& K^{+}_{\theta_{n} \;\theta_{n}} = \left[ F Y \dot{T} \right]_{\Sigma}, \label{ecib}  \\
&&  K^{-}_{t \; t} =  0, \\
&& K^{-}_{\theta_{n}\; \theta_{n}} = \left[\frac{R R'}{e ^\lambda}
\right]_{\Sigma}.\label{ecic}
\end{eqnarray}
With the help of Eqs.~(\ref{fjc}) - ~(\ref{ecic}) and
~(\ref{eq:fe}) , the total energy entrapped within the surface
$\Sigma$ can be given by \cite{cm}
\begin{equation}
M = \frac{n-1}{2} {\mathcal{M}(r)}.\label{sm}
\end{equation}
Thus, the junction conditions demand that the HD Schwarzschild
mass $M$ is given by Eq.~(\ref{sm}).

\section{Horizons}
The apparent horizon is formed when the boundary of trapped $n$
spheres are formed. In spherical dust collapse, the event horizon
coincides with the apparent horizon at the boundary of the
spherical mass distribution. The apparent horizon is the solution
of
\begin{equation}
g^{ab} R_{,a} R_{,b} = - \dot{R}^2 + f(r) + 1 = 0.
\end{equation}
Upon using Eqs.~(\ref{eq:el}) and ~(\ref{eq:fe}), we have
\begin{equation}
\Lambda R^{n+1} - \frac{n(n+1)}{2}R^{n-1} + \frac{n(n+1)}{2}
\mathcal{M} = 0.\label{eq:ah}
\end{equation}
For $\Lambda = 0$ we have the Schwarzschild horizon $R^{n-1} =
\mathcal{M}$, and for $\mathcal{M} = 0$ we have the de Sitter
horizon $R = \pm \sqrt {n(n+1)/2 \; \Lambda}$.
The approximate solutions of Eq.~(\ref{eq:ah})to first order are
\begin{eqnarray}
R_{{bh}}& =& R^{(0)}_{{bh}} + R^{(1)}_{{bh}} + \;.\;.\;. \nonumber\\
& = &\mathcal{M}^{1/n-1} + \frac{2\Lambda}{n(n-1)(n+1)}
\mathcal{M}^{3/n-1}.\;.\;.,\label{b}
\end{eqnarray}
where $R_{{bh}}$ is the radius of the black hole event horizon.
\begin{eqnarray}
R_{{ch}}& =& R^{(0)}_{{ch}} + R^{(1)}_{{ch}} + \;.\;.\;. \nonumber\\
& =& \left[\frac{n(n+1)}{2\Lambda}\right]^{1/2} - \frac{1}{2}
\left[\frac{2\Lambda}{n(n+1)}\right]^{n/2}
\mathcal{M}.\;.\;.,\label{c}
\end{eqnarray}
where $R_{{ch}}$ is the radius of the cosmological event horizon
\cite{xd}. There exist a critical solution of Eq.~(\ref{eq:ah}),
where two roots coincides and there is only one horizon. The time
for the formation of apparent horizon, from Eq.~(\ref{eq:R}) is
\begin{equation}
t_{AH} = t_{0}(r) - \sqrt{\frac{2n}{(n+1)\Lambda}}\;\sinh^{-1}
\left[\sqrt{\frac{2\Lambda}{n(n+1)}}\;\mathcal{M}^{1/n-1}\right].
\end{equation}
In the limit as $\Lambda \rightarrow
0$, using Eq.~(\ref{TBR}), we obtain:
\begin{equation}
t_{AH} = t_{0}(r) - \frac{2}{n+1} \mathcal{M}^{1/n-1}.\label{AH}
\end{equation}
Equation (\ref{AH}), gives the time for the formation of event
horizon in HD Tolman-Bondi-de Sitter space-time. A necessary
condition for the singularity to be globally naked is $t_{AH} \geq
t_{0}(0)$.
\section{Causal structure  of singularities}
It has been shown \cite{rn} that Shell-crossing singularities are
characterized by $R'=0$ and $R>0$. On the other hand the
singularity at $R=0$ is, where all matter shells collapses to a
zero physical radius and hence known as shell focussing
singularity.

We shall consider the case $t \geq t_{0}$. In the context of the
Tolman-Bondi models the shell crossings are defined to be surfaces
on which $R' = 0$ ($R > 0$) and where the density $\zeta$
diverges. A regular extremum in $R$ along constant time slices may
occur without causing a shell crossing, provided $\zeta(t,r)$ does
not diverge. By Eq.~(\ref{eq:edt}), this implies $\mathcal{M}' =
0$ where ever $R' = 0$ and also that the surface $R' = 0$ remain
at fixed $R$. Now Eq.~(\ref{eq:RD}) implies $t_{o}' = 0$. Thus the
condition for a regular maximum in $R(t,r)$ is that $\mathcal{M}'
= 0$, $t_{o}' = 0$ hold at the same $R$. It has been shown
\cite{rn} that shell crossing singularities are gravitationally
weak and hence such singularities cannot be considered seriously.

Next, we turn our attention to shell-focusing singularities.
Christodoulou \cite{dc} pointed out in the 4D case that the
non-central singularities are not naked. Hence, we shall confine
our discussion to the central shell focusing singularity. The
energy density diverge at $t=t_{0}(r)$ indicating the presence of
curvature singularity \cite{he}. It is known that, depending upon
the inhomogeneity factor, the 4D Tolman-Bondi solutions admits a
central shell focusing naked singularity in the sense that
outgoing geodesics emanate from the singularity. Here we wish to
investigate the similar situation in our HD Tolman-Bondi-de Sitter
space-time. We consider a class of models such that
\begin{subequations}
\label{ss}
\begin{eqnarray}
&& \mathcal{M}(r) = \gamma r^{n-1}, \label{ssa} \\
&& t_{0}(r) = B r. \label{ssc}
\end{eqnarray}
\end{subequations}
This class of models for 4D space-time is discussed in \cite{kl,
jl, op}. The parameter $B$ gives the inhomogeneity of the
collapse. For $B=0$ all shells collapse at the same time.  For
higher $B$ the outer shells collapse much later than the central
shell. We are interested in the causal structure of the space-time
when the central shell collapses to the center ($R=0$). From
Eqs.~(\ref{eq:edt}) and ~(\ref{ssa}), the energy density at the
singularity
\begin{equation}
\zeta = \frac{n(n-1) \gamma}{2r^2}, \label{eds1}
\end{equation}
and the equation of general density becomes
\begin{widetext}
\begin{equation}
\zeta = \frac{\Lambda}{\sinh^{2}{\left[\sqrt{\frac{(n+1)\Lambda}{2n}}\frac{(B
- y)}{y}\;t\right]} \left[1 +
\sqrt{\frac{2(n+1)\Lambda}{n}}\;\frac{t}{y}\;\frac{B}{n-1}\;\coth
{\left[\sqrt{\frac{(n+1)\Lambda}{2n}}\;\frac{t}{y}\;(B - y)\right]}\right]}.
 \label{ed}
\end{equation}
\end{widetext}
As $t \rightarrow t_{0}(r)$ i.e. in approach to
singularity, we have $\sinh\alpha \approx \alpha$ and $\coth
\alpha \approx 1/\alpha$ then Eq.~(\ref{ed}) reduces to:
\begin{equation}
\zeta = \frac{n(n-1) y^2}{(n+1) (B-y) \left[\frac{n+1}{2}B -
\frac{n-1}{2}y\right]t^2} = \frac{C(y)}{t^2}. \label{eds2}
\end{equation}
The nature (a naked singularity or a black hole) of the
singularity can be characterized by the existence of radial null
geodesics emerging from the singularity. The singularity is at
least locally naked if there exist such geodesics, and if no such
geodesics exist, it is a black hole. The critical direction is the
Cauchy horizon. This is the first outgoing null geodesic emanating
from $r=t=0$. The Cauchy horizon of the space-time has $y = t/r$ =
const \cite{lz,jl}. The equation for outgoing null geodesics is
\begin{eqnarray}
\frac{dt}{dr} = R'.
\end{eqnarray}
Hence along the Cauchy horizon, we have
\begin{eqnarray}
R' = y, \label{ch}
\end{eqnarray}
and using Eqs.~(\ref{ch}) and ~(\ref{eq:RD}), with our choice of
the scale,
 we obtain the following algebraic equation:
\begin{eqnarray}
y \left(1 - \frac{y}{B}\right)^{\frac{n-1}{n+1}} = \left(\frac{n+1}{2}
B \sqrt{\gamma}\right)^{\frac{2}{n+1}}\left[1 - \frac{n-1}{n+1}\frac{y}{B}\right].\label{eq:ae}
\end{eqnarray}
To facilitate comparison with Ghosh and Beesham \cite{gb}, we
introduce a a new relation between $B$ and $\gamma$ as,
\begin{equation}
B =\frac{2}{n+1} \frac{1}{\sqrt{\gamma}},
\end{equation}
with the help of above relation and after some rearrangement, Eq.
~(\ref{eq:ae}) takes the exactly the same form as in Ghosh and
Beesham \cite{gb}
\begin{equation}
y\left[1 -
\frac{n+1}{2}\;\sqrt{\gamma}\;y\right]^{\frac{n-1}{n+1}} +
\frac{n-1}{2}\;\sqrt{\gamma}\;y - 1 = 0. \label{eq:ae1}
\end{equation}
This algebraic equation governs the behavior of the tangent vector
near the  singular point. The central shell focusing singularity
is at least locally naked, if Eq.~(\ref{eq:ae1}) admits one or
more positive real roots. Hence in the absence of positive real
roots, the collapse will always lead to a black hole. Since the
Eq.~(\ref{eq:ae1}) is the same as that obtained by Ghosh and
Beesham \cite{gb} when the metric is asymptotically flat i.e.,
$\Lambda = 0$. Consequently the values of roots for the geodetic
tangent and the condition for these values to be real and positive
are the same as those obtained for the asymptotically flat
situation in \cite{gb}. As a result, the Tolman-Bondi-de Sitter
space-time has same singularity behavior as the Tolman-Bondi
space-time in both 4D and HD.  Thus the results of collapsing
inhomogeneous dust in de Sitter background are similar to that of
collapsing inhomogeneous dust in Minkowskian background, as it
should have been expected, since when $t \rightarrow t_{0}(r)$ the
cosmological term ${2 \Lambda R^2}/{n(n + 1)}$ is negligible.  It
is known that the Tolman-Bondi metric ($\Lambda=0$), in the 4D
case is extensively used for studying the formation of naked
singularities in spherical gravitational collapse. It has been
found that Tolman-Bondi metric admit both naked singularities and
black holes form depending upon the choice of initial data.
Indeed, both analytical \cite{dj}-\cite{jd1} and numerical results
\cite{es} in dust indicate the critical behavior governing the
formation of black holes or naked singularities. One can now
safely assert that end state of 4D Tolman-Bondi collapse is now
completely known in dependence of choice of initial data. A
similar situation also occurs in HD Tolman-Bondi collapse
\cite{gb,bsc} for a class of model under discussion. Since the
presence of $\Lambda>0$ does not change the final fate of
Inhomogeneous dust collapse. Hence, we can conclude both naked
singularities and black holes can form in the HD spherical
inhomogeneous dust collapse with a positive $\Lambda$.

Further, it is shown that, in the class models discussed here, the
formation of a black hole is facilitated with introduction of the
extra dimensions or, in other words, the naked singularity
spectrum gets continuously covered in higher dimensional
space-times. This is valid in the in null fluid collapse \cite{gd}
and  scalar field collapse \cite{kmy} in higher dimensional
space-times. Thus it appears that the singularity will be
completely covered for very large dimensions of the space-time.
Recently, It has been shown that cosmic censorship can be restored
in some classes of inhomogeneous dust gravitational collapse
models when space-time dimension to be $N \geq 6$, i.e., the naked
singularities of this model can be removed by going to higher
dimensions \cite{gj}. However, this is valid with only smooth
initial profiles.
\section{Concluding remarks}
In this paper, We have shown that the 4D spherically symmetric
solution describing inhomogeneous dust collapse with a
cosmological term go over to $(n+2)$-dimensional spherically
symmetric solution and essentially retaining its physical behavior
and when $n=2$, one recovers the 4D Tolman-Bondi-de Sitter
solutions \cite{djcj,sg,cf}. Thus we have obtained Tolman-Bondi-de
Sitter metric in arbitrary dimensions and the junction condition
for static and non-static space-times are deduced. We have also
utilized this solution to study the end state of collapsing star
and showed that there exists a regular initial data which leads to
a naked singularity.

Our purpose was to investigate the collapse of inhomogeneous dust
shells in an expanding de Sitter background, to find out if the
naked singularity occurs in this situation and to compare any
difference with the similar collapse in the asymptotically flat
case.   We have obtained a condition for the occurrence of a naked
singularity in the collapse of dust shells in an expanding
background which is the same as that obtained when the background
is asymptotically flat. This very fact establishes that the
space-time is asymptotically flat or not does not make any
difference to the occurrence of a naked singularity. This is
evident at least the in the class of models defined by
Eq.~(\ref{ss}).

Penrose \cite{rp1} has conjectured that it seems unlikely that a
$\Lambda$-term will really make much difference to the singularity
structure in a collapse. The relevance of $\Lambda$ is really only
at the cosmological scale.  Our results are consistent with this.
However, The introduction of a cosmological constant changes
scenario in many ways. There are now several apparent horizons
instead of one. However, only two apparent horizon are physical,
namely the black hole horizon and the cosmological horizon.  Other
results derived in \cite{cf} do carry over to HD space-time
essentially with same physical behavior hence not presented to
avoid duplication.

Finally, the result obtained would also be relevant in the context
of superstring theory which is often said to be next "theory of
everything", and for an interpretation of how critical behaviour
depends on the dimensionality of the space-time.

\noindent {\bf Acknowledgment:} The authors would like to thank
the IUCAA, Pune for kind hospitality while part of this work was
being done.   One of the author(SGG) would like to thank Director,
BITS Pilani, Dubai for continuous encouragements.

\noindent
\end{document}